\documentclass[11pt,a4paper]{article}
\usepackage{float}
\usepackage{authblk}
\usepackage{placeins}
\usepackage[T1]{fontenc}
\usepackage[utf8]{inputenc}
\usepackage{lmodern}
\usepackage{amsmath,amssymb}
\usepackage{booktabs}
\usepackage{graphicx}
\usepackage{tikz}
\usetikzlibrary{arrows.meta,decorations.pathmorphing,positioning,decorations.markings}
\tikzset{
  fermion/.style = {postaction={decorate},
    decoration={markings, mark=at position 0.5 with {\arrow{>}}}, line width=0.8pt},
  boson/.style   = {decorate, decoration={snake, amplitude=1.3pt, segment length=6pt}, line width=0.8pt},
  ext/.style     = {font=\normalsize}
}
\usepackage{subcaption}
\usepackage[numbers,sort&compress]{natbib}
\DeclareUnicodeCharacter{2032}{\ensuremath{^{\prime}}}
\DeclareUnicodeCharacter{2033}{\ensuremath{^{\prime\prime}}}
\DeclareUnicodeCharacter{2019}{'}
\DeclareUnicodeCharacter{2018}{'}
\DeclareUnicodeCharacter{2013}{-}
\DeclareUnicodeCharacter{2212}{-}
\usepackage[colorlinks=true,linkcolor=blue,citecolor=blue,urlcolor=blue]{hyperref}

\title{Discovery Prospects for a Leptophilic Gauge Boson $Z_\ell$ at CEPC and ILC}
\author{S. O. Kara}
\affil{\centering Niğde Ömer Halisdemir University, Bor Vocational School, 51240, Niğde, Türkiye}
\date{}
\begin{document}
\maketitle

\begin{abstract}
We investigate the discovery prospects of a leptophilic gauge boson $Z_\ell$ at 
future $e^+e^-$ colliders, focusing on a comparative study of the 
Circular Electron--Positron Collider (CEPC) and the International Linear Collider (ILC). 
Such a state naturally arises from an additional $U(1)'_\ell$ gauge symmetry, 
under which quarks remain neutral while all leptons carry a universal charge, 
motivated by neutrino oscillations and scenarios of physics beyond the Standard Model (SM). 
As a clean benchmark, we study the process $e^+e^- \to \mu^+\mu^-$, including 
realistic effects of initial-state radiation (ISR) and beamstrahlung (BS). 
Our results indicate that CEPC, with its very high luminosity at $\sqrt{s}=240$~GeV, 
can probe couplings down to $g_\ell \approx 10^{-3}$ for $Z_\ell$ masses up to about 220~GeV, 
while the ILC extends the sensitivity to heavier states in the multi-hundred GeV range 
through its higher $\sqrt{s}$ stages. These findings demonstrate the strong 
complementarity of circular and linear colliders in exploring purely leptophilic interactions.
\end{abstract}


\section{Introduction}

The Standard Model (SM) of particle physics has been remarkably successful 
in describing the fundamental constituents of matter and their interactions~\cite{1}. 
Nevertheless, the observation of neutrino oscillations~\cite{2,3} 
and the violation of individual lepton flavor numbers clearly point towards new 
physics beyond the SM. Among the simplest and most intriguing possibilities is 
the existence of an additional abelian gauge symmetry under which quarks remain 
neutral while all leptons carry a universal charge. Such a symmetry is particularly 
well motivated by the fact that neutrino oscillations already require new dynamics 
in the lepton sector, suggesting that leptons may play a privileged role in 
extensions of the SM. In addition, the $U(1)'_\ell$ construction is anomaly-free 
when all lepton families are assigned identical charges, providing a theoretically 
consistent and minimal extension of the gauge structure. The associated gauge boson, 
conventionally denoted as the leptophilic photon $Z_\ell$, couples exclusively 
to leptons and thus realizes a genuinely leptophilic interaction~\cite{4,5,6,7,8,9,10}.

Electron–positron colliders provide an ideal environment to probe this kind of 
new physics due to their clean experimental conditions and well-defined initial states. 
In particular, the upcoming Circular Electron–Positron Collider (CEPC)~\cite{11} 
and the International Linear Collider (ILC)~\cite{12} represent two complementary 
approaches for the next generation of lepton colliders. While the CEPC aims at extremely 
high luminosity around the Higgs factory energy, the ILC offers the flexibility of 
operating at both $\sqrt{s}=250$~GeV and $500$~GeV, thereby extending the reach 
into the TeV regime. This complementarity renders them powerful and mutually reinforcing 
facilities to explore new leptophilic interactions.

Beyond the general motivation, it is worth stressing that CEPC and ILC 
offer complementary advantages that have not yet been systematically 
compared in the leptophilic context. Earlier works have focused either on 
generic $Z^\prime$ scenarios including quark couplings or on simplified 
parton-level estimates for purely leptophilic interactions~\cite{7,8}, 
without incorporating realistic collider and detector effects. 
By contrast, the present study includes both initial-state radiation (ISR) and 
beamstrahlung (BS) as well as acceptance cuts, thereby providing sensitivity 
projections that are directly applicable to the experimental programs 
of CEPC and ILC.

Moreover, the comparison is timely in view of the global strategy for 
future colliders. CEPC has been highlighted as a central project in China, 
designed to operate as a Higgs factory with unprecedented luminosity, 
while the ILC has received strong international attention as a linear 
machine with higher center-of-mass energy options. Establishing the 
relative strengths of these colliders in probing leptophilic gauge 
interactions is therefore not only of theoretical interest but also of 
practical importance for optimizing the physics reach of upcoming 
facilities. This complementarity is particularly relevant since 
hadron colliders have very limited sensitivity to leptophilic states, 
making lepton machines unique laboratories for such searches.

Taken together, these considerations underline both the novelty and the 
significance of the present analysis. Our results provide the first 
direct, side-by-side comparison of CEPC and ILC capabilities for 
discovering a leptophilic photon, and thus contribute to the broader 
discussion of physics opportunities at next-generation lepton colliders.

In this work, we present a comparative phenomenological study of the discovery 
potential for the leptophilic photon $Z_\ell$ at CEPC and ILC, focusing on the 
clean channel $e^+e^- \to \mu^+\mu^-$. We evaluate the expected sensitivities by 
incorporating realistic effects of ISR, BS, and detector acceptance. 
This channel provides a distinctive signature with minimal SM background 
and therefore constitutes an optimal probe of purely leptophilic interactions. 
In this way, our analysis extends previous parton-level estimates~\cite{7,8} 
and offers a more reliable assessment of the capabilities of future $e^+e^-$ colliders.

\section{Theoretical Framework}

A minimal and economical way to extend the Standard Model (SM) is to introduce 
an additional abelian gauge symmetry $U(1)'_\ell$ under which all leptons carry 
a universal charge, while quarks remain neutral. This idea dates back to early 
studies of generalized gauge structures~\cite{13,14,15} 
and has been revisited in various contexts of leptophilic or leptonic interactions
~\cite{8,9,16}. The new gauge boson, denoted 
$Z_\ell$, couples exclusively to leptons and therefore realizes a genuinely 
leptophilic interaction.

The covariant derivative is modified to include the new gauge interaction,
\begin{equation}
D_\mu = \partial_\mu + i g'_\ell q_\ell Z'_\mu ,
\end{equation}
where $g'_\ell$ is the gauge coupling constant, $q_\ell$ the universal lepton 
charge, and $Z'_\mu$ the new gauge field. The relevant terms of the Lagrangian read
\begin{equation}
\mathcal{L} \supset -\frac{1}{4} F'_{\mu\nu} F'^{\mu\nu} 
+ \bar{\ell} \, i \gamma^\mu D_\mu \ell 
+ (D_\mu \Phi)^\dagger (D^\mu \Phi) - V(\Phi),
\end{equation}
where $\Phi$ is a scalar field charged under $U(1)'_\ell$ whose vacuum expectation 
value (vev) spontaneously breaks the new symmetry, thereby generating a mass for 
the leptophilic boson $Z_\ell$.

The interaction between $Z_\ell$ and the leptonic current can be expressed as
\begin{equation}
\mathcal{L}_{\text{int}} = g_\ell Z_{\ell,\mu} J^\mu_{\text{lep}}, 
\qquad J^\mu_{\text{lep}} = \sum_{\ell=e,\mu,\tau} \bar{\ell}\gamma^\mu \ell ,
\label{eq:lep_current}
\end{equation}
which highlights the universal coupling of $Z_\ell$ to all charged leptons. A crucial theoretical requirement is the absence of gauge anomalies, which would 
otherwise spoil the consistency of the theory at the quantum level~\cite{17,18,19,20}. 
In the present setup, anomalies cancel if all three lepton families are assigned 
the same $U(1)'_\ell$ charge. With this choice, the $[SU(2)_W]^2 U(1)'_\ell$, 
$[U(1)_Y]^2 U(1)'_\ell$, and $[U(1)'_\ell]^3$ anomalies vanish.

Mixing effects can in principle arise in two ways. First, kinetic mixing with 
the hypercharge gauge boson $B_\mu$ may occur through
\begin{equation}
\mathcal{L}_{\text{mix}} = - \frac{\epsilon}{2} F^{Y}_{\mu\nu} F'^{\mu\nu} ,
\end{equation}
where $\epsilon$ parametrizes the mixing strength. Such terms are typically 
induced at loop level~\cite{14}, but precision electroweak data require 
$\epsilon \lesssim 10^{-2}$~\cite{7}. Recent works~\cite{21} 
have studied electroweak precision constraints in nearly-degenerate $Z'$–$Z$ systems, 
showing that even small kinetic and mass mixing generate observable shifts in the 
oblique parameters $S,T,U$, consistent with existing data. Second, after electroweak 
symmetry breaking, the SM $Z$ boson and the new $Z_\ell$ can mix via the scalar sector. 
The mass-squared matrix in the $(Z, Z_\ell)$ basis is
\begin{equation}
M^2 = 
\begin{pmatrix}
M_Z^2 & \delta M^2 \\
\delta M^2 & M_{Z_\ell}^2
\end{pmatrix},
\end{equation}
with $\delta M^2$ induced by the vev of $\Phi$. Further studies~\cite{22} 
computed corrections to electroweak observables from such mixings in the context 
of anomalies like the $W$-boson mass shift, finding constraints that closely match 
the $\theta \lesssim 10^{-3}$ limit. In the parameter space of interest here, both 
kinetic and mass mixing are negligible, so $Z_\ell$ couples purely to leptons at 
tree level. Model-independent analyses~\cite{23} further support the treatment 
of mixing effects as subdominant for the benchmark values considered here.

Constraints from precision electroweak measurements imply a lower bound on the ratio 
between the new boson mass and its coupling,
\begin{equation}
\frac{M_{Z_\ell}}{g_\ell} \gtrsim 7~\text{TeV}.
\label{eq:bound}
\end{equation}
Representative exclusion limits on $g_\ell$ for benchmark masses are summarized 
in Table~\ref{tab:limits}. Electroweak precision data require 
$g_\ell \lesssim 1.4\times 10^{-2}$ at $M_{Z_\ell}\sim 100~\text{GeV}$, with the 
bound gradually relaxing to ${\cal O}(10^{-1})$ in the multi-TeV regime.

\begin{table}[tbp]
\centering
\caption{95\% C.L. exclusion limits on the leptophilic coupling $g_\ell$ from electroweak precision fits.}
\label{tab:limits}
\begin{tabular}{cc}
\toprule
$M_{Z_\ell}$ [GeV] & Limit on $g_\ell$ \\
\midrule
100   & $< 1.4\times 10^{-2}$ \\
240   & $< 3.4\times 10^{-2}$ \\
250   & $< 3.6\times 10^{-2}$ \\
500   & $< 7.1\times 10^{-2}$ \\
1000  & $< 1.4\times 10^{-1}$ \\
3000  & $< 4.3\times 10^{-1}$ \\
\bottomrule
\end{tabular}
\end{table}

Altogether, the framework is anomaly-free, predictive, and renormalizable. 
The $Z_\ell$ couples universally to leptons, has negligible couplings to quarks, 
and only tiny mixing with SM gauge bosons. These features make future 
high-luminosity lepton colliders the ideal environment to search for $Z_\ell$, 
as hadronic machines are comparatively much less sensitive.

\section{Collider Setup}

To assess the sensitivity to leptophilic interactions at future $e^+e^-$ machines, 
we base our analysis on the baseline beam parameters summarized in 
Table~\ref{tab:collider}. For CEPC we adopt the design values reported in the 
Conceptual Design Report (2018)~\cite{11}, and for the ILC we follow the 
Technical Design Report (2013)~\cite{12}; these choices are consistent with 
Ref.~\cite{24}. Throughout, $\sigma_x$ and $\sigma_y$ denote the horizontal 
and vertical root-mean-square (rms) beam sizes at the interaction point (IP), 
$\sigma_z$ the longitudinal bunch length (also an rms size, in $\mu$m), $N_b$ 
the particles per bunch (in units of $10^{10}$), and $L$ the integrated luminosity 
accumulated over the data-taking period. The CEPC is envisaged as a circular Higgs 
factory operating at $\sqrt{s}=240~\text{GeV}$ with a target integrated luminosity 
of about $6~\text{ab}^{-1}$, providing a clean environment with very high statistics 
for precision Higgs and electroweak studies. In contrast, the ILC is a linear collider 
foreseen to begin at $\sqrt{s}=250~\text{GeV}$ ($1.35~\text{ab}^{-1}$) with an upgrade 
path to $\sqrt{s}=500~\text{GeV}$ ($1.8~\text{ab}^{-1}$)~\cite{12}, thereby extending 
the direct kinematic reach for heavier states. Although other proposed lepton colliders 
such as FCC-ee~\cite{25} and CLIC~\cite{26} offer additional benchmark scenarios, they 
lie beyond the scope of the present CPC-focused comparison; hence, in what follows we 
restrict our quantitative analysis to CEPC and ILC.

\begin{table}[tbp]
\centering
\caption{Baseline beam parameters adopted for CEPC and ILC. Integrated luminosities are given in ab$^{-1}$.}
\label{tab:collider}
\begin{tabular}{lcccccc}
\toprule
Collider & $\sqrt{s}$ (GeV) & $L$ (ab$^{-1}$) & $\sigma_x$ (nm) & $\sigma_y$ (nm) & $\sigma_z$ ($\mu$m) & $N_b$ ($\times 10^{10}$) \\
\midrule
CEPC    & 240 & 6.0  & 16.0  & 0.12 & 4.4   & 3.7 \\
ILC-250 & 250 & 1.35 & 515.0 & 7.7  & 300.0 & 2.0 \\
ILC-500 & 500 & 1.80 & 474.0 & 5.9  & 300.0 & 2.0 \\
\bottomrule
\end{tabular}
\end{table}

In order to obtain realistic sensitivity estimates, we include the effects of 
initial-state radiation (ISR) and beamstrahlung (BS), which modify the effective 
luminosity spectrum. ISR is implemented following the structure-function approach 
of Skrzypek and Jadach~\cite{27,28}, while BS is modeled using 
the official collider beam spectra provided in the design reports~\cite{29}. 
These effects reduce the effective collision energy and broaden the invariant-mass 
distributions, thereby affecting both total and differential cross sections. 
Throughout our study, we simulate the signal process $e^+e^- \to \mu^+\mu^-$ via 
$\gamma/Z/Z_\ell$ exchange and the corresponding Standard Model background using the CalcHEP framework~\cite{30,31,32}.
 Detector effects are approximated by applying 
acceptance cuts of $|\cos\theta_\mu| < 0.95$ and assuming high muon identification 
efficiency consistent with design studies.

\section{Signal and Background Analysis}

The signal process considered is $e^+e^- \to \mu^+\mu^-$, mediated by $Z_\ell$. This channel is particularly advantageous because the muon 
final state can be reconstructed with high efficiency and excellent momentum 
resolution, making it the cleanest probe of leptophilic interactions.

\begin{figure}[t]
  \centering

  \begin{subfigure}[t]{0.45\textwidth}
    \centering
    \begin{tikzpicture}[x=1cm,y=1cm]
      \coordinate (e-)  at (-1.3,  0.9);
      \coordinate (e+)  at (-1.3, -0.9);
      \coordinate (v1)  at ( 0.0,  0.0);
      \coordinate (v2)  at ( 2.6,  0.0);
      \coordinate (mu-) at ( 3.9,  0.9);
      \coordinate (mu+) at ( 3.9, -0.9);

      \draw[fermion] (e-) -- (v1) node[midway, above, ext] {$e^-$};
      \draw[fermion] (e+) -- (v1) node[midway, below, ext] {$e^+$};
      \draw[boson]   (v1) -- (v2) node[midway, above, ext] {$Z_\ell$};
      \draw[fermion] (v2) -- (mu-) node[midway, above, ext] {$\mu^-$};
      \draw[fermion] (v2) -- (mu+) node[midway, below, ext] {$\mu^+$};
    \end{tikzpicture}
    \caption{Signal: $e^+e^- \to Z_\ell \to \mu^+\mu^-$}
  \end{subfigure}\hfill
  %
  \begin{subfigure}[t]{0.45\textwidth}
    \centering
    \begin{tikzpicture}[x=1cm,y=1cm]
      \coordinate (e-)  at (-1.3,  0.9);
      \coordinate (e+)  at (-1.3, -0.9);
      \coordinate (v1)  at ( 0.0,  0.0);
      \coordinate (v2)  at ( 2.6,  0.0);
      \coordinate (mu-) at ( 3.9,  0.9);
      \coordinate (mu+) at ( 3.9, -0.9);

      \draw[fermion] (e-) -- (v1) node[midway, above, ext] {$e^-$};
      \draw[fermion] (e+) -- (v1) node[midway, below, ext] {$e^+$};
      \draw[boson]   (v1) -- (v2) node[midway, above, ext] {$\gamma,\,Z$};
      \draw[fermion] (v2) -- (mu-) node[midway, above, ext] {$\mu^-$};
      \draw[fermion] (v2) -- (mu+) node[midway, below, ext] {$\mu^+$};
    \end{tikzpicture}
    \caption{Background: $e^+e^- \to \gamma^*/Z \to \mu^+\mu^-$}
  \end{subfigure}

  \caption{Representative Feynman diagrams for the signal process 
  $e^+e^- \to Z_\ell \to \mu^+\mu^-$ (left) and the dominant Standard Model 
  background $e^+e^- \to \gamma^*/Z \to \mu^+\mu^-$ (right).}

  \label{fig:feynman}
\end{figure}
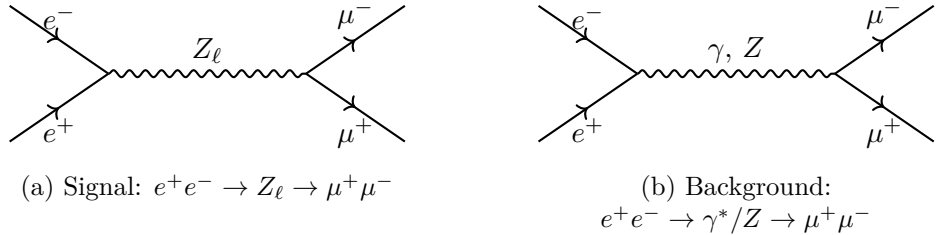

The dominant irreducible background arises from the Standard Model Drell--Yan 
production via photon and $Z$ exchange, $e^+e^- \to \gamma^*/Z \to \mu^+\mu^-$. 
This background is theoretically well understood and experimentally clean, 
allowing deviations from the SM prediction to be directly attributed to the presence of a $Z_\ell$ contribution~\cite{33}. Reducible backgrounds from processes 
such as $e^+e^- \to \tau^+\tau^-$ or multihadronic final states are negligible 
once standard selection criteria are applied. For this reason the di-muon 
channel provides a uniquely sensitive and robust avenue to search for a 
leptophilic photon at future lepton colliders. The signal and background processes are illustrated in Fig.~\ref{fig:feynman}.

In our analysis, ISR and beamstrahlung effects are fully included, as discussed 
in the previous section. Detector-level effects are approximated by applying 
acceptance cuts of $|\cos\theta_\mu| < 0.95$ and $p_T^\mu > 10~\text{GeV}$, 
consistent with the design reports of CEPC and ILC. After cuts, the SM background 
is reduced while the $Z_\ell$ signal retains a significant fraction of events, 
enhancing sensitivity in the high-mass tails of distributions.

The statistical significance is evaluated using a binned likelihood method 
based on the invariant-mass spectrum. For each benchmark mass hypothesis, 
we derive exclusion limits on the coupling $g_\ell$ at 95\% C.L.~\cite{34,35} 
the predicted signal-plus-background spectrum with the background-only 
expectation. These results are presented in the following section.

\begin{figure}[tbp]
  \centering
  \includegraphics[width=0.75\textwidth]{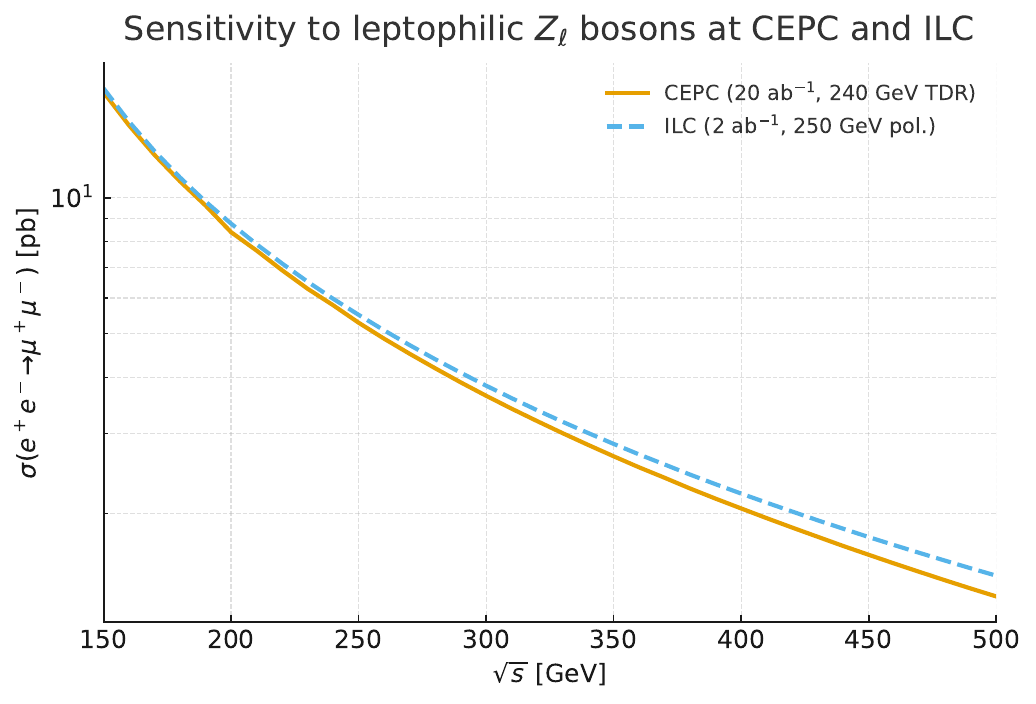}
  \caption{Total cross section $\sigma(e^+e^-\!\to\mu^+\mu^-)$ as a function of the center-of-mass energy $\sqrt{s}$ for CEPC and ILC benchmark configurations. 
  The CEPC baseline at $\sqrt{s}=240~\mathrm{GeV}$ with $20~\mathrm{ab}^{-1}$ integrated luminosity provides excellent sensitivity to light leptophilic gauge bosons, 
  while the higher-energy ILC stages (250 and 500~GeV) extend the accessible mass range to heavier $Z_\ell$ states.}
  \label{fig:sigmavsenergy}
\end{figure}

The energy dependence of the total cross section is displayed in 
Fig.~\ref{fig:sigmavsenergy}. At CEPC, operating at $\sqrt{s}=240$~GeV, 
the high luminosity allows excellent sensitivity to deviations from the 
Standard Model prediction, while the ILC at $\sqrt{s}=250$ and $500$~GeV 
extends the accessible range in energy and probes heavier $Z_\ell$ states. 
The comparison illustrates the complementarity of circular and linear 
machines in exploring leptophilic interactions.

\begin{figure}[H]
  \centering

  \begin{subfigure}[t]{1.0\textwidth}
    \centering
    \includegraphics[width=\linewidth]{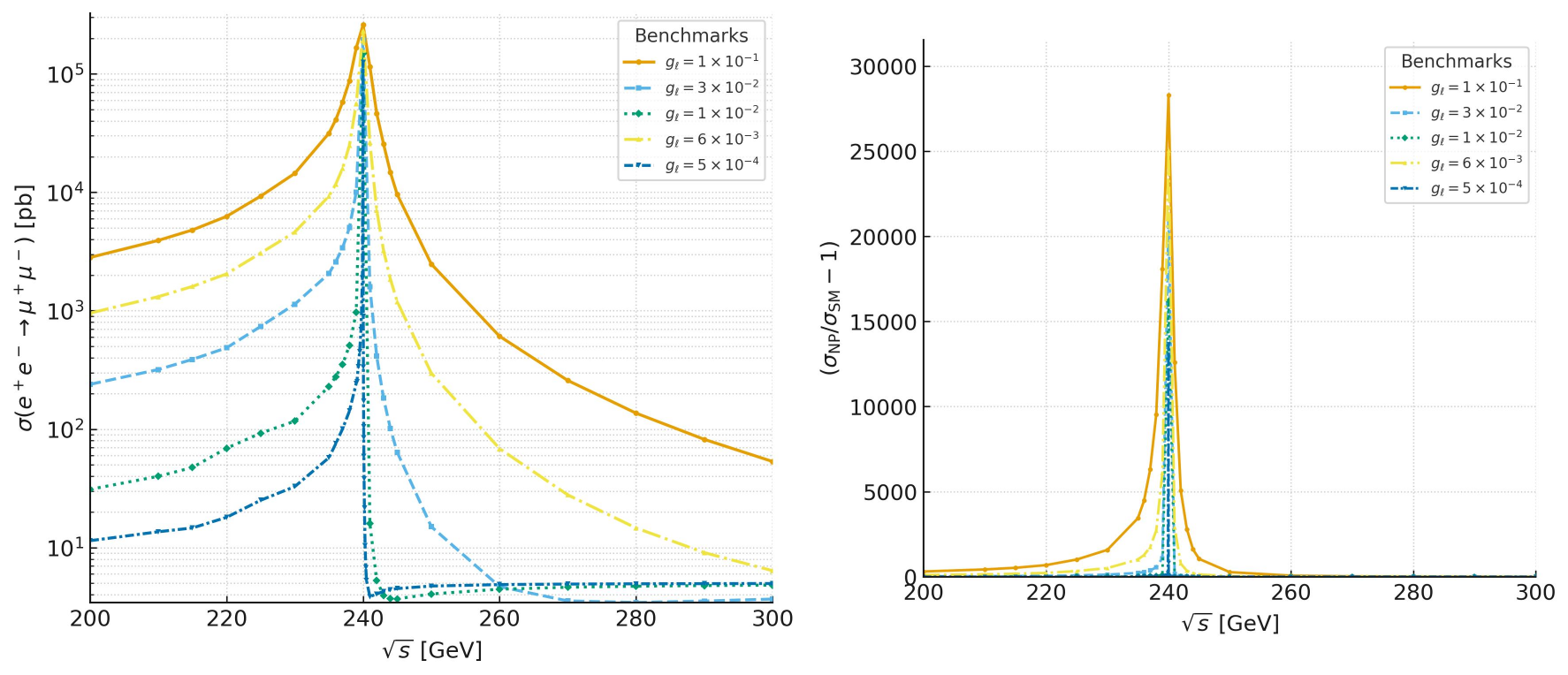}
    \caption{CEPC, $\sqrt{s}=240$ GeV}
    \label{fig:cepc_sigma}
  \end{subfigure}

  \vspace{2mm}

  \begin{subfigure}[t]{0.75\textwidth}
    \centering
    \includegraphics[width=\linewidth]{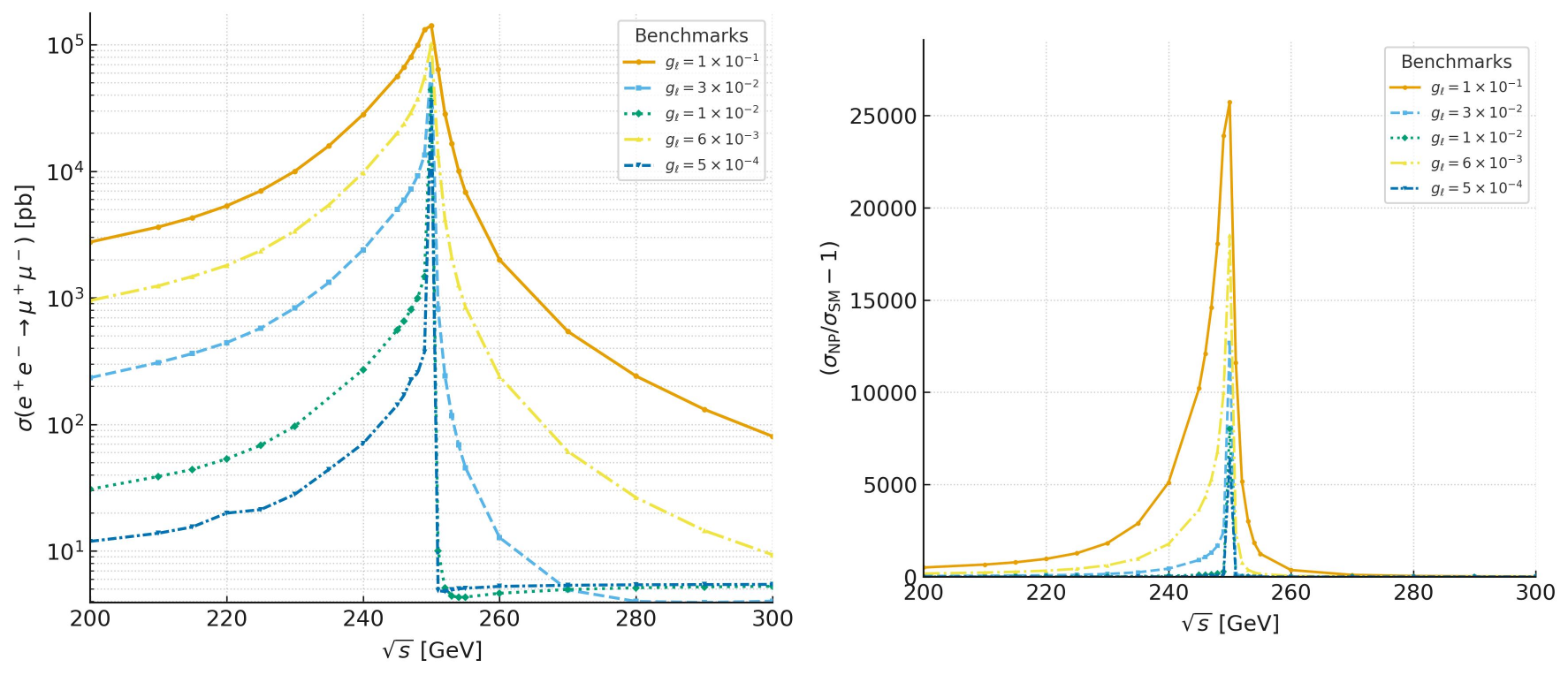}
    \caption{ILC, $\sqrt{s}=250$ GeV}
    \label{fig:ilc250_sigma}
  \end{subfigure}\hfill
  \begin{subfigure}[t]{0.75\textwidth}
    \centering
    \includegraphics[width=\linewidth]{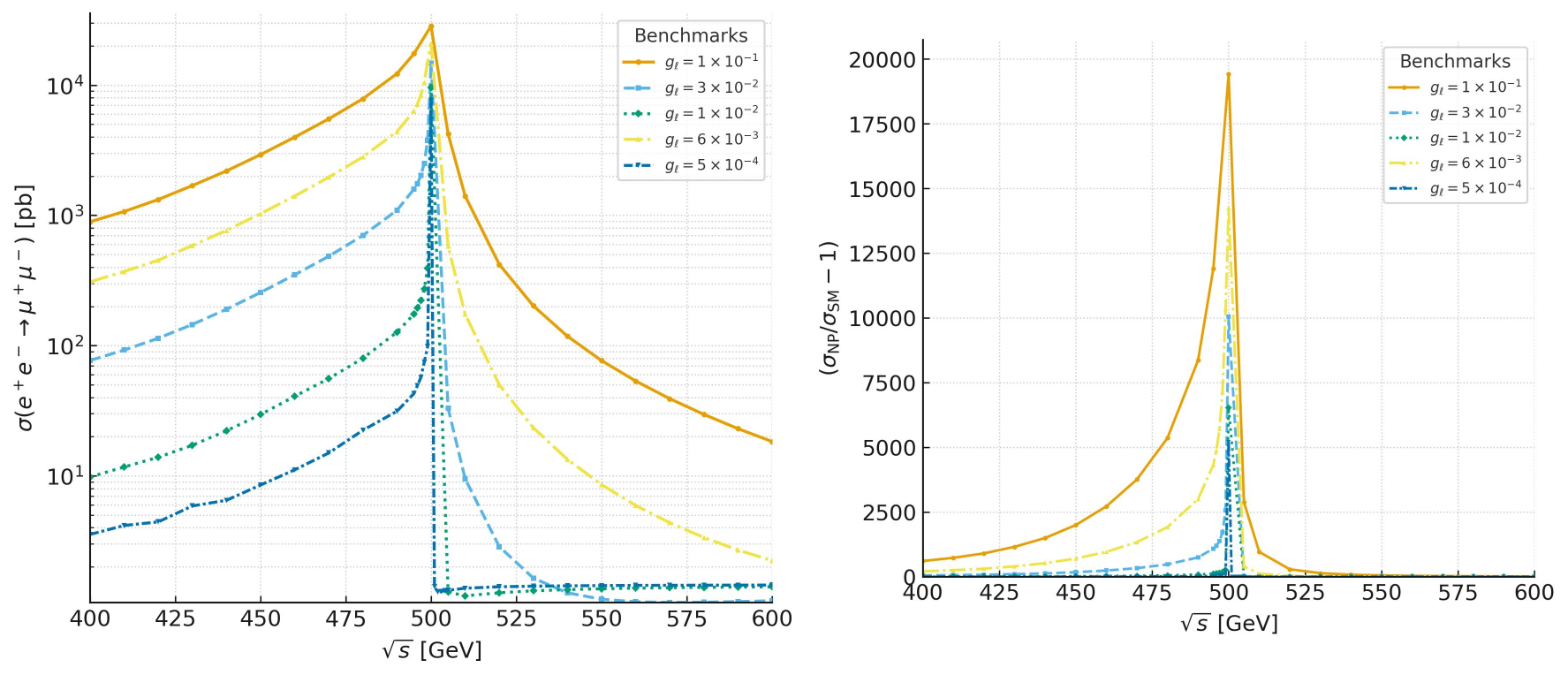}
    \caption{ILC, $\sqrt{s}=500$ GeV}
    \label{fig:ilc500_sigma}
  \end{subfigure}

  \caption{Total cross section $\sigma(e^+e^-\!\to\mu^+\mu^-)$ and its relative deviation $(\sigma_{\rm NP}/\sigma_{\rm SM}-1)$ as functions of $M_{Z_\ell}$ for representative couplings $g_\ell$.
  (a) CEPC at $\sqrt{s}=240~\text{GeV}$ with $20~\text{ab}^{-1}$ integrated luminosity (TDR baseline).
  (b) ILC at $\sqrt{s}=250~\text{GeV}$ with $2~\text{ab}^{-1}$ polarized beams.
  (c) ILC at $\sqrt{s}=500~\text{GeV}$, where higher energies extend the sensitivity to heavier $Z_\ell$ states.
  Right-hand panels show the fractional deviation from the SM prediction, highlighting the discovery potential even for small couplings.}
  \label{fig:sigma_vs_mzl}
\end{figure}

In Fig.~\ref{fig:sigma_vs_mzl}, we show the total cross section as well as the relative deviation $(\sigma_{\rm NP}/\sigma_{\rm SM}-1)$ as functions of the new boson mass $M_{Z_\ell}$ for representative values of the coupling $g_\ell$. 
The CEPC panel demonstrates that even small couplings can induce visible deviations thanks to its very high luminosity, although the accessible mass range is limited by the lower center-of-mass energy. 
The ILC panels illustrate how higher energies broaden the discovery potential, with the $500~\mathrm{GeV}$ setup being particularly effective in probing multi-hundred-GeV $Z_\ell$ states. 
Overall, these complementary results emphasize the excellent sensitivity of future $e^+e^-$ colliders to leptophilic gauge interactions.

\begin{figure}[H]
  \centering

  \begin{subfigure}[t]{0.85\textwidth}
    \centering
    \includegraphics[width=\linewidth]{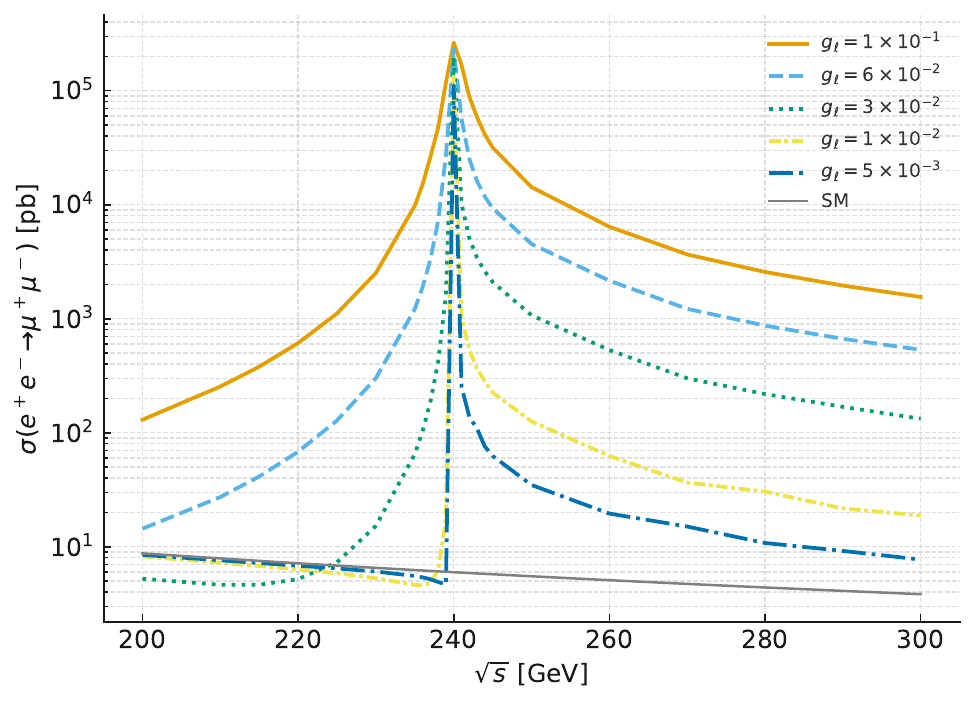}
    \caption{CEPC, $\sqrt{s}=240$ GeV}
    \label{fig:cepc_limit}
  \end{subfigure}

  \vspace{2mm}

  \begin{subfigure}[t]{0.48\textwidth}
    \centering
    \includegraphics[width=\linewidth]{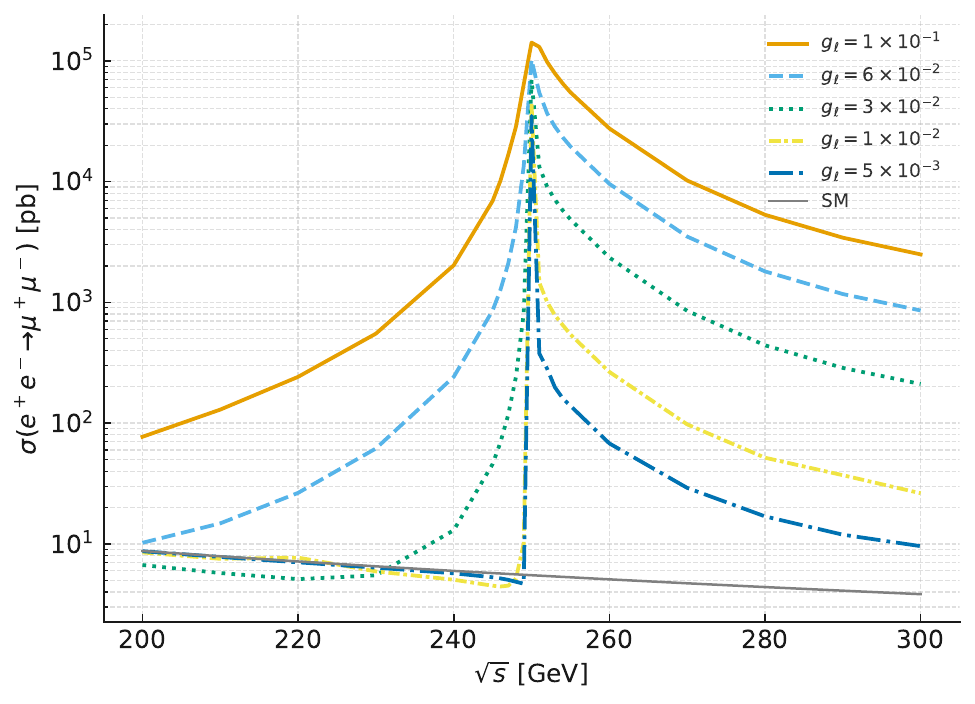}
    \caption{ILC, $\sqrt{s}=250$ GeV}
    \label{fig:ilc250_limit}
  \end{subfigure}\hfill
  \begin{subfigure}[t]{0.48\textwidth}
    \centering
    \includegraphics[width=\linewidth]{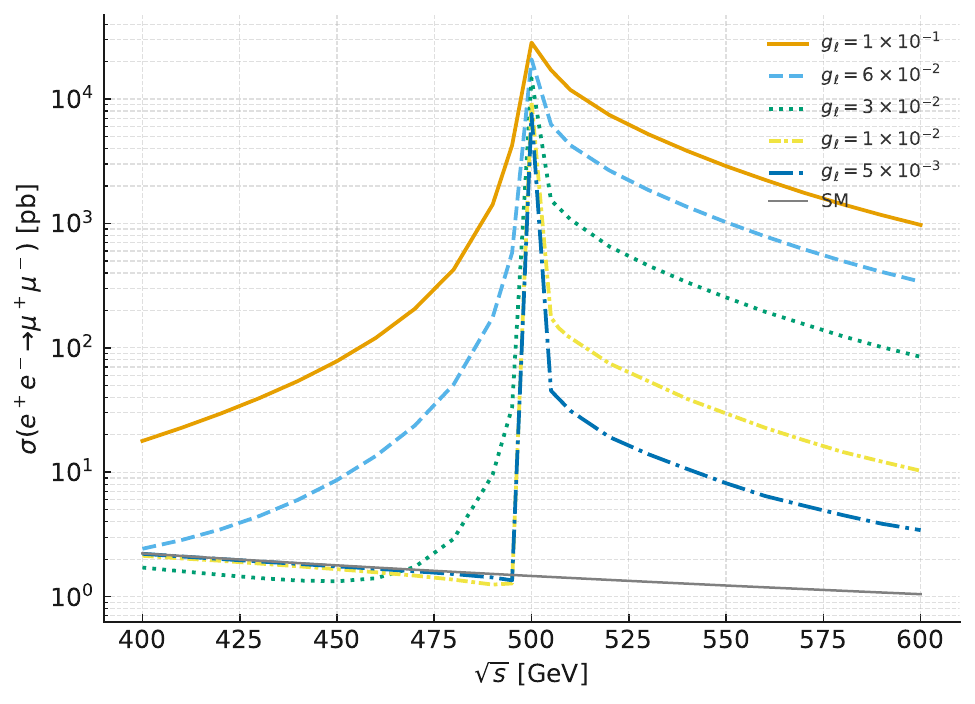}
    \caption{ILC, $\sqrt{s}=500$ GeV}
    \label{fig:ilc500_limit}
  \end{subfigure}

  \caption{Energy scans of the total cross section $\sigma(e^+e^-\!\to\mu^+\mu^-)$ for representative couplings $g_\ell$ at future $e^+e^-$ colliders.
(a)~CEPC at $\sqrt{s}=240~\mathrm{GeV}$ (20~$\mathrm{ab}^{-1}$, TDR baseline).
(b)~ILC at $\sqrt{s}=250~\mathrm{GeV}$ (2~$\mathrm{ab}^{-1}$, polarized).
(c)~ILC at $\sqrt{s}=500~\mathrm{GeV}$.
Line styles distinguish the benchmark scenarios; the SM expectation is shown in gray.}
  \label{fig:limits}
\end{figure}

The dependence of the total cross section on the center-of-mass energy is 
illustrated in Fig.~\ref{fig:limits}. The CEPC projection (Fig.~\ref{fig:cepc_limit}) shows excellent sensitivity near 
$\sqrt{s}=240$~GeV due to its very high luminosity, while the ILC panels 
(Figs.~\ref{fig:ilc250_limit} and~\ref{fig:ilc500_limit}) extend the reach 
to higher energies. In particular, the 500~GeV configuration enhances the 
prospects for discovering a heavy $Z_\ell$ by probing deviations from the 
Standard Model spectrum well into the TeV scale.

\begin{figure}[H]
  \centering
  \begin{subfigure}[t]{0.85\textwidth}
    \centering
    \includegraphics[width=\linewidth]{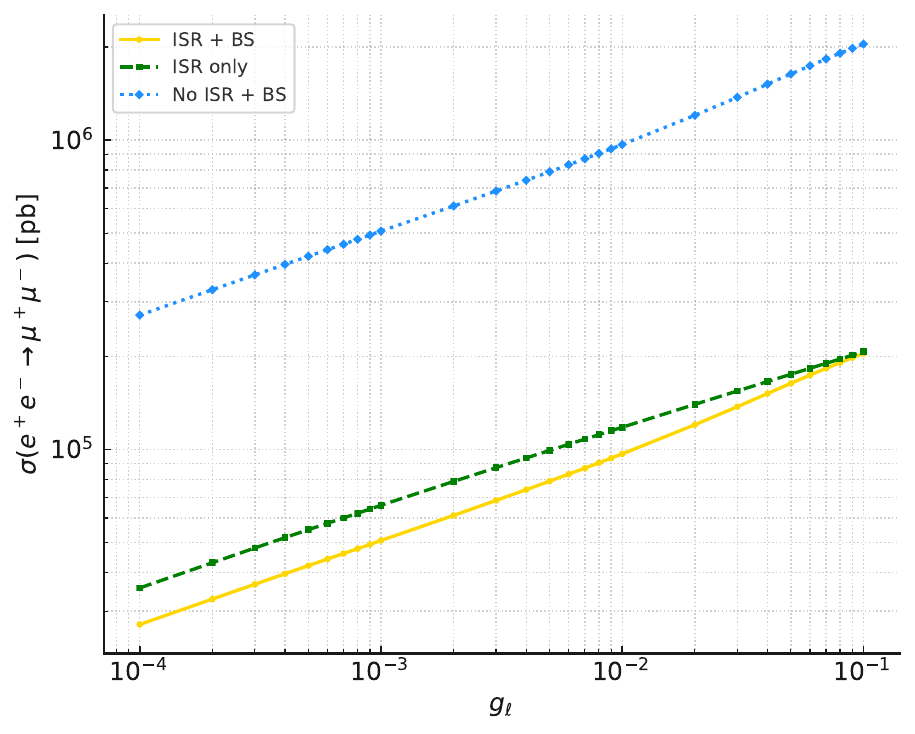}
    \caption{CEPC, $\sqrt{s}=240$ GeV}
    \label{fig:cepc_gl}
  \end{subfigure}

  \vspace{1.5mm}

  \begin{subfigure}[t]{0.48\textwidth}
    \centering
    \includegraphics[width=\linewidth]{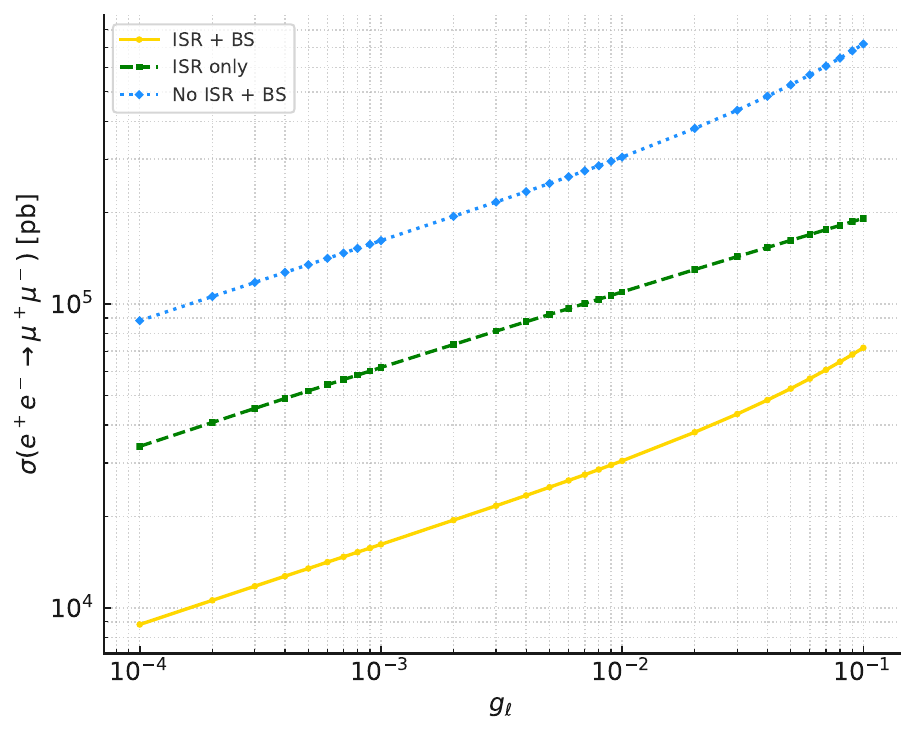}
    \caption{ILC, $\sqrt{s}=250$ GeV}
    \label{fig:ilc250_gl}
  \end{subfigure}\hfill
  \begin{subfigure}[t]{0.48\textwidth}
    \centering
    \includegraphics[width=\linewidth]{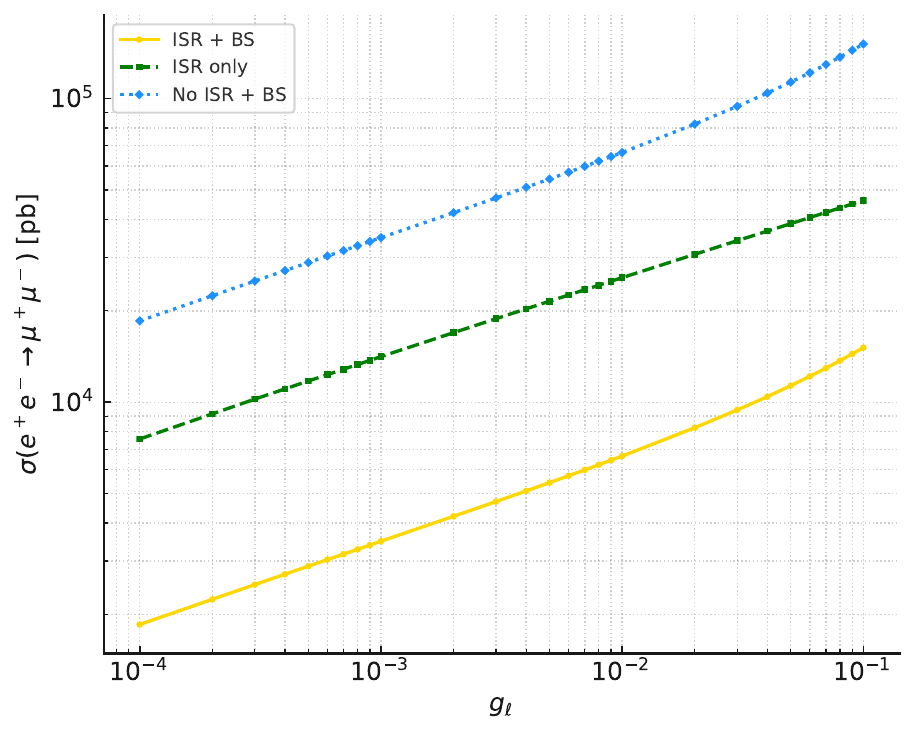}
    \caption{ILC, $\sqrt{s}=500$ GeV}
    \label{fig:ilc500_gl}
  \end{subfigure}

  \caption{Total cross section $\sigma(e^+e^-\!\to\mu^+\mu^-)$ 
  as a function of the leptophilic coupling $g_\ell$ for CEPC and ILC setups, 
  illustrating the impact of initial-state radiation (ISR) and beamstrahlung (BS). 
  (a)~CEPC at $\sqrt{s}=240~\mathrm{GeV}$, 
  (b)~ILC at $\sqrt{s}=250~\mathrm{GeV}$, 
  and (c)~ILC at $\sqrt{s}=500~\mathrm{GeV}$. 
  Solid, dotted, and dash–dot lines correspond to the full ISR+BS simulation, 
  ISR only, and the idealized No ISR+BS case, respectively. 
  Both axes are displayed on logarithmic scales.}
  \label{fig:gl_dependence}
\end{figure}

The dependence of the total cross section on the leptophilic coupling $g_\ell$
is displayed in Fig.~\ref{fig:gl_dependence}.  
For each collider setup, three configurations are compared:  
the idealized case without ISR/BS, the ISR–only case, and the full ISR+BS simulation.  
At the CEPC ($\sqrt{s}=240~\mathrm{GeV}$), beamstrahlung has only a mild effect,
while ISR dominates the suppression of the effective cross section.
For the ILC, both the 250 and 500~GeV options
(Figs.~\ref{fig:ilc250_gl} and~\ref{fig:ilc500_gl})
exhibit a stronger sensitivity to ISR+BS, particularly at larger $g_\ell$ values.
These comparisons clearly demonstrate the necessity of including realistic
beam effects when estimating discovery sensitivities at future $e^+e^-$ colliders.

\begin{figure}[H]
  \centering

  \begin{subfigure}[t]{0.85\textwidth}
    \centering
    \includegraphics[width=\linewidth]{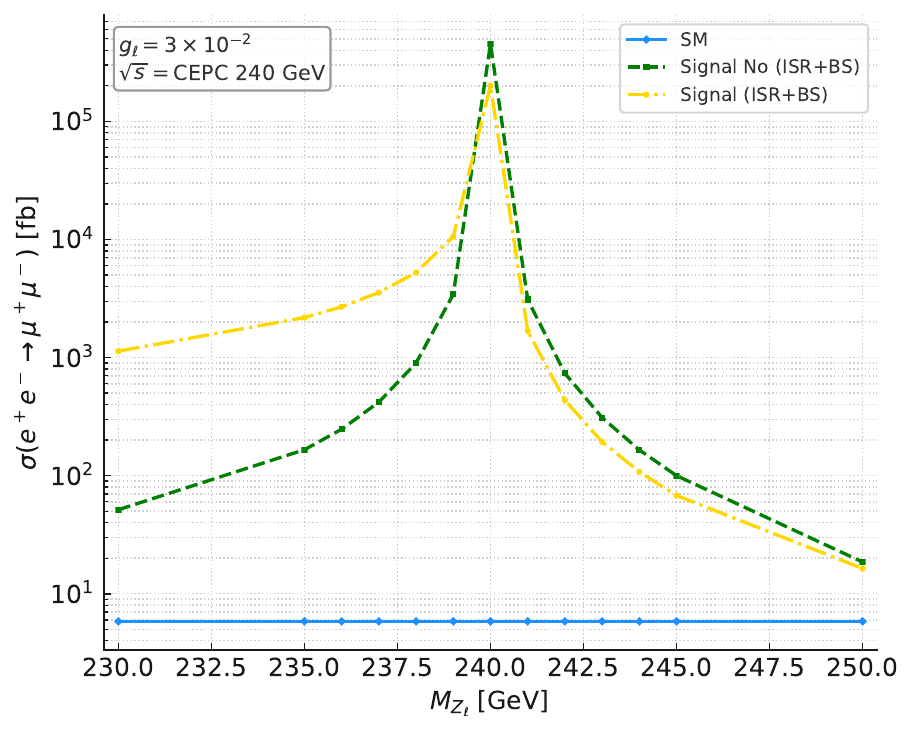}
    \caption{CEPC, $\sqrt{s}=240$ GeV}
    \label{fig:cepc_isr_bs}
  \end{subfigure}

  \vspace{2mm}

  \begin{subfigure}[t]{0.48\textwidth}
    \centering
    \includegraphics[width=\linewidth]{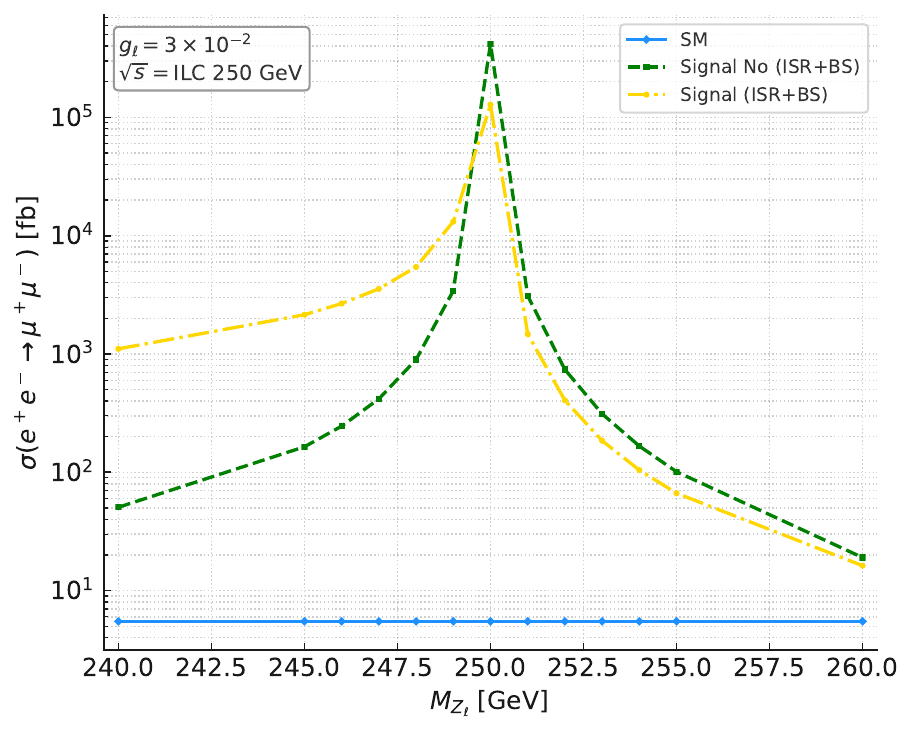}
    \caption{ILC, $\sqrt{s}=250$ GeV}
    \label{fig:ilc250_isr_bs}
  \end{subfigure}\hfill
  \begin{subfigure}[t]{0.48\textwidth}
    \centering
    \includegraphics[width=\linewidth]{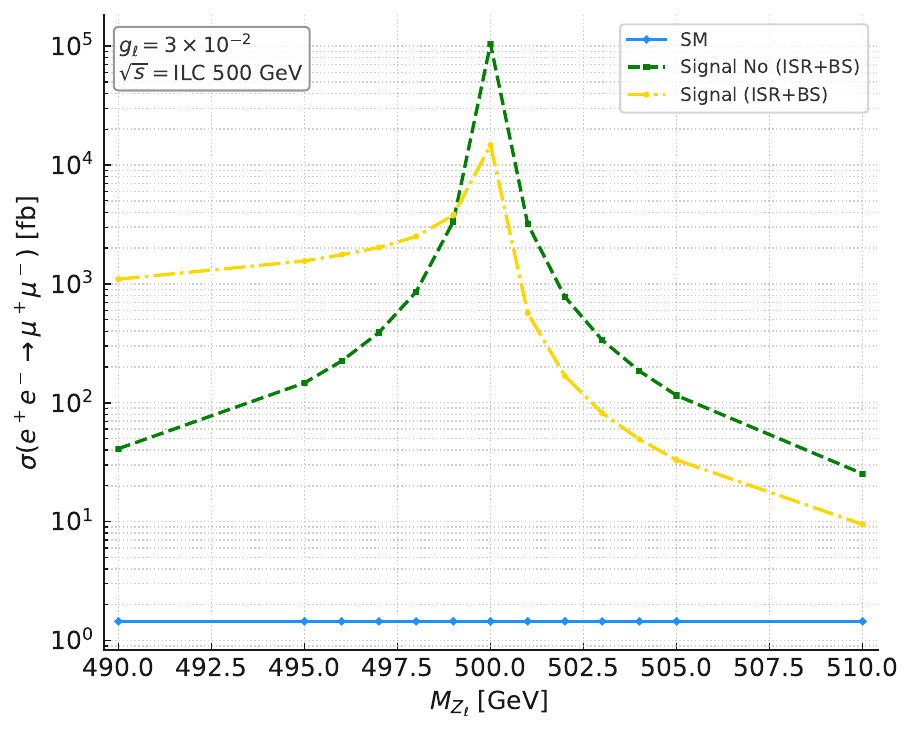}
    \caption{ILC, $\sqrt{s}=500$ GeV}
    \label{fig:ilc500_isr_bs}
  \end{subfigure}

  \caption{Cross section $\sigma(e^+e^- \!\to\! \mu^+\mu^-)$ as a function of 
  $M_{Z_\ell}$ for $g_\ell=3\times10^{-2}$. Comparison shown for SM background, 
  signal without ISR/BS, and full ISR+BS. CEPC peak remains narrow, ILC-250 broadens, 
  and ILC-500 widens further though deviation from SM is still distinct.}
  \label{fig:isr_bs}
\end{figure}

The impact of ISR and beamstrahlung is illustrated in Fig.~\ref{fig:isr_bs}, 
where the cross section is plotted as a function of $M_{Z_\ell}$ for 
$g_\ell=3\times10^{-2}$. At CEPC (Fig.~\ref{fig:cepc_isr_bs}) the resonance 
appears as a narrow peak, while at ILC-250 (Fig.~\ref{fig:ilc250_isr_bs}) it is 
moderately broadened, and at ILC-500 (Fig.~\ref{fig:ilc500_isr_bs}) the peak 
becomes significantly wider due to stronger beam effects. In all cases, however, 
the deviation from the SM background remains visible, showing that ISR and 
beamstrahlung reduce but do not eliminate the discovery potential~\cite{36,37}.

\begin{figure}[H]
  \centering

  \begin{subfigure}[t]{0.85\textwidth}
    \centering
    \includegraphics[width=\linewidth]{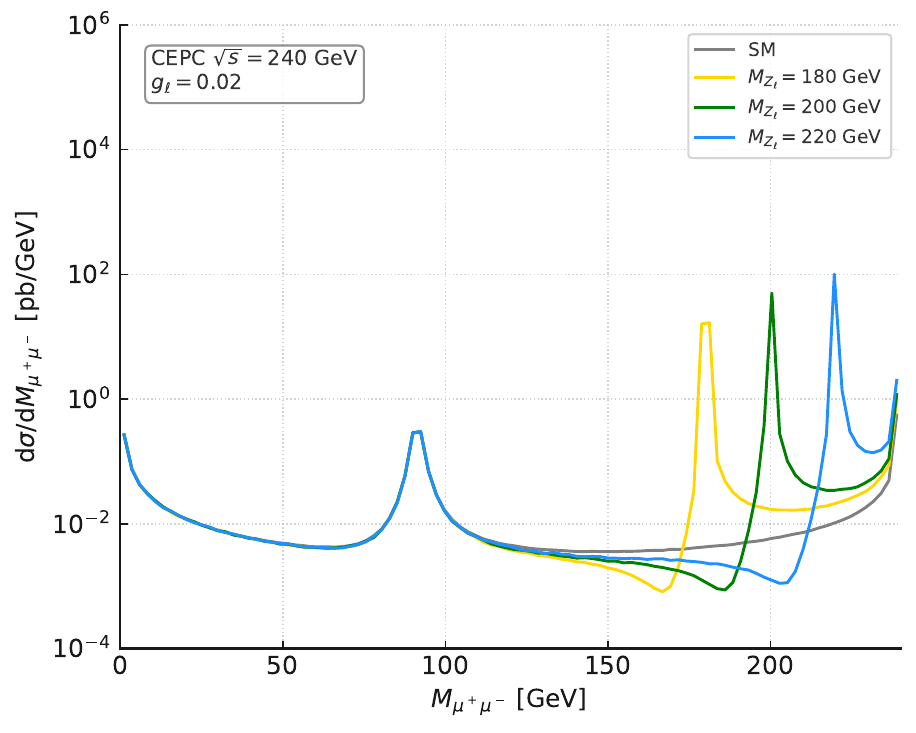}
    \caption{CEPC, $\sqrt{s}=240$ GeV}
    \label{fig:cepc_invmass}
  \end{subfigure}

  \vspace{2mm}

  \begin{subfigure}[t]{0.48\textwidth}
    \centering
    \includegraphics[width=\linewidth]{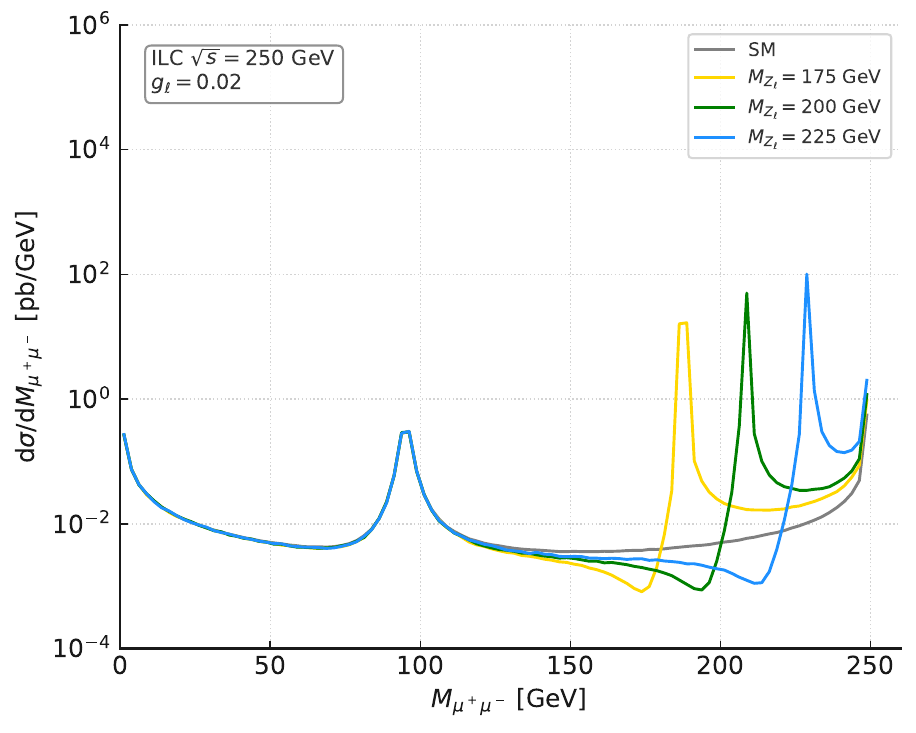}
    \caption{ILC, $\sqrt{s}=250$ GeV}
    \label{fig:ilc250_invmass}
  \end{subfigure}\hfill
  \begin{subfigure}[t]{0.51\textwidth}
    \centering
    \includegraphics[width=\linewidth]{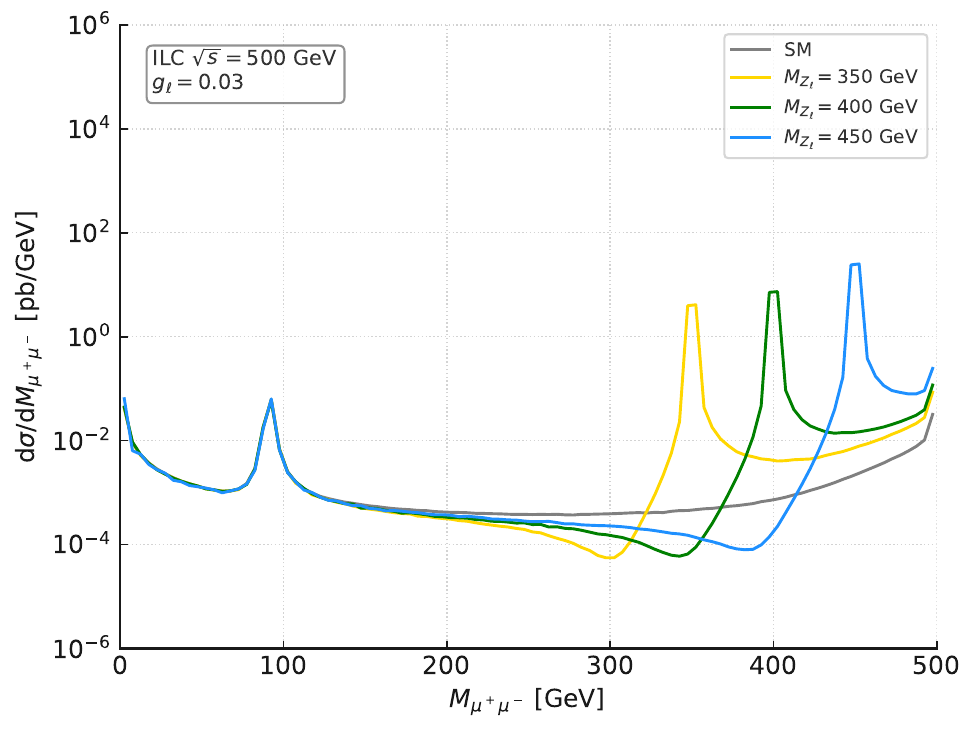}
    \caption{ILC, $\sqrt{s}=500$ GeV}
    \label{fig:ilc500_invmass}
  \end{subfigure}

  \caption{Invariant mass distributions of the di-muon system for representative 
$Z_\ell$ benchmark masses at future $e^+e^-$ colliders. 
(a)~CEPC at $\sqrt{s}=240~\mathrm{GeV}$ with $g_\ell=0.02$ shows pronounced peaks in the 180--220~GeV range. 
(b)~ILC at $\sqrt{s}=250~\mathrm{GeV}$ displays similar structures near 200~GeV, 
and (c)~ILC at $\sqrt{s}=500~\mathrm{GeV}$ with $g_\ell=0.03$ extends the sensitivity 
to resonances up to about 450~GeV. 
The narrow $Z_\ell$ resonances stand out clearly above the smooth Standard Model background, 
illustrating the discovery potential and excellent mass resolution achievable 
at future lepton colliders.}

  \label{fig:invmass}
\end{figure}

The invariant mass distribution of the di-muon pair provides a striking signature of a leptophilic resonance. 
Figure~\ref{fig:invmass} shows the expected spectra for several representative $Z_\ell$ masses. 
At the CEPC ($\sqrt{s}=240~\mathrm{GeV}$, Fig.~\ref{fig:cepc_invmass}), narrow peaks emerge around 180--220~GeV for $g_\ell=0.02$, 
clearly separated from the smooth Standard-Model background. 
The ILC-250 configuration ($\sqrt{s}=250~\mathrm{GeV}$, Fig.~\ref{fig:ilc250_invmass}) exhibits similar features near 200~GeV, 
while the ILC-500 setup ($\sqrt{s}=500~\mathrm{GeV}$, Fig.~\ref{fig:ilc500_invmass}) with $g_\ell=0.03$ 
demonstrates sensitivity to heavier resonances up to about 450~GeV. 
These distributions highlight the excellent mass resolution and discovery reach achievable at future $e^+e^-$ colliders, 
with visible $Z_\ell$ peaks even for moderate couplings.

\section{Results and Discussion}

Before presenting the numerical projections, it is important to verify that the
benchmark scenarios considered here are consistent with existing experimental limits.
Current collider searches place only mild constraints on purely leptophilic gauge
bosons, since hadron colliders are mostly sensitive to quark–coupled interactions.
Dedicated dilepton resonance searches at the LHC~\cite{ATLAS:2019erb,CMS:2021ctt}
typically assume universal couplings to both quarks and leptons, which are absent
in our model. Consequently, direct LHC limits can be substantially relaxed, allowing
sub–TeV $Z_\ell$ masses for couplings below $g_\ell \lesssim 0.1$.
Additional constraints from LEP-II contact–interaction analyses~\cite{LEP:2003aa}
and low–energy precision data~\cite{BaBar:2014zli,MuonG-2:2021ojo}
are also satisfied across the parameter space explored in this study.
Therefore, all benchmark points used in the following analysis remain consistent
with current bounds and provide a valid basis for discovery projections.

\medskip

Having established the signal and background framework, we now turn to the 
projected sensitivities and discovery reaches for the leptophilic photon 
$Z_\ell$ at CEPC and ILC.

\begin{figure}[H]
  \centering

  \begin{subfigure}[t]{0.85\textwidth}
    \centering
    \includegraphics[width=\linewidth]{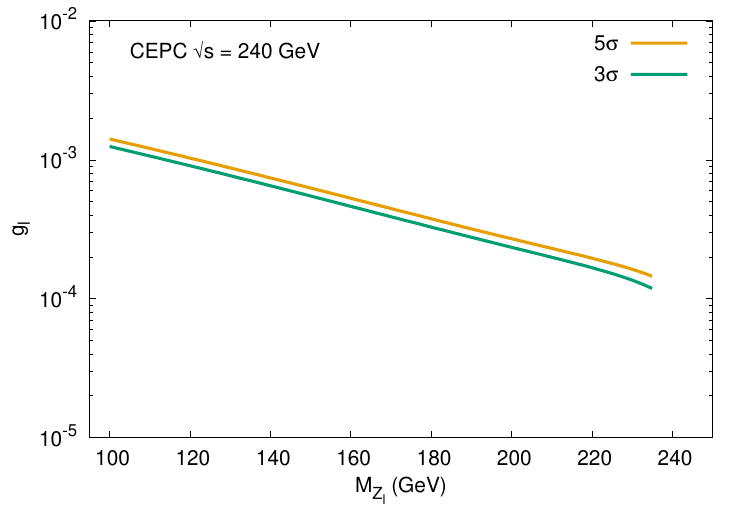}
    \caption{CEPC, $\sqrt{s}=240$ GeV}
    \label{fig:cepc_disc}
  \end{subfigure}

  \vspace{2mm}

  \begin{subfigure}[t]{0.48\textwidth}
    \centering
    \includegraphics[width=\linewidth]{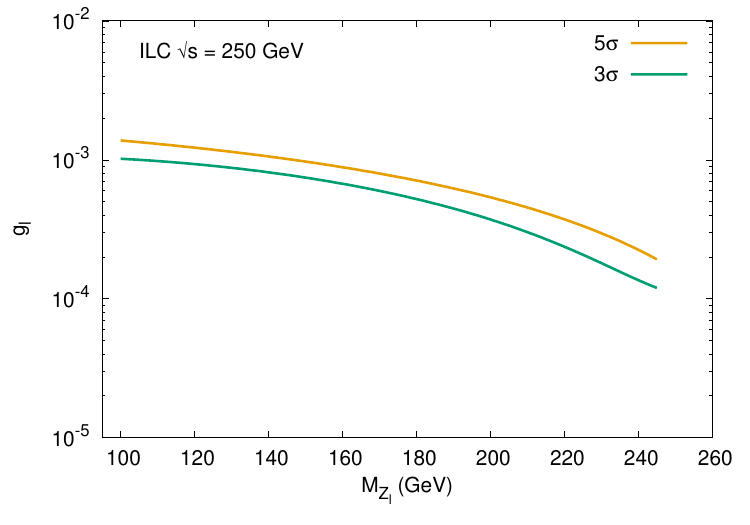}
    \caption{ILC, $\sqrt{s}=250$ GeV}
    \label{fig:ilc250_disc}
  \end{subfigure}\hfill
  \begin{subfigure}[t]{0.48\textwidth}
    \centering
    \includegraphics[width=\linewidth]{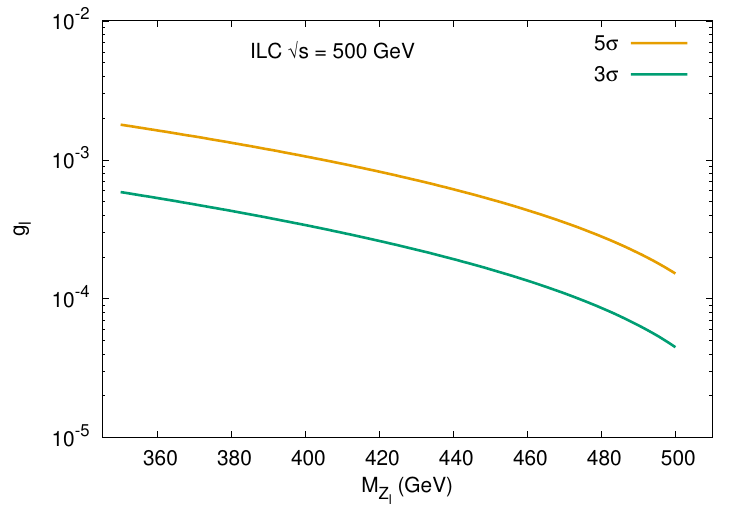}
    \caption{ILC, $\sqrt{s}=500$ GeV}
    \label{fig:ilc500_disc}
  \end{subfigure}

  \caption{Projected $3\sigma$ and $5\sigma$ discovery reaches in the $(M_{Z_\ell}, g_\ell)$ plane.
(a)~CEPC at $\sqrt{s}=240~\mathrm{GeV}$ demonstrates sensitivity down to $g_\ell\simeq10^{-3}$ for $M_{Z_\ell}\lesssim220~\mathrm{GeV}$.
(b)~ILC-250 extends the mass range while maintaining comparable coupling reach, 
and (c)~ILC-500 further improves the mass coverage up to the multi-hundred-GeV region.
The curves correspond to statistical significances computed with full ISR+BS simulation,
highlighting the complementarity of circular and linear $e^+e^-$ colliders.}

  \label{fig:disc_reach}
\end{figure}

The discovery potential of future $e^+e^-$ colliders is summarized in 
Fig.~\ref{fig:disc_reach}, which displays the 3$\sigma$ and 5$\sigma$ 
significance contours in the $(M_{Z_\ell},\,g_\ell)$ plane~\cite{25,38,39,40}. 
At CEPC (Fig.~\ref{fig:cepc_disc}), couplings down to $g_\ell \approx 10^{-3}$ 
are accessible for $M_{Z_\ell}$ up to about 220~GeV. The ILC-250 projection 
(Fig.~\ref{fig:ilc250_disc}) extends the reach in mass while maintaining 
similar sensitivity to small couplings, and the ILC-500 setup 
(Fig.~\ref{fig:ilc500_disc}) pushes the accessible region further into 
the TeV scale. These results emphasize the complementarity of circular 
and linear colliders in exploring the leptophilic gauge boson scenario.

\medskip

In addition to the total cross section, we have also examined differential observables 
that are sensitive to angular effects, such as the forward–backward asymmetry 
$A_{\mathrm{FB}} = (\sigma_F - \sigma_B)/(\sigma_F + \sigma_B)$, where 
$\sigma_F$ and $\sigma_B$ denote the forward and backward cross sections, respectively. 
The interference between the SM and leptophilic amplitudes induces characteristic 
deviations in $A_{\mathrm{FB}}$, particularly near the $Z_\ell$ resonance. 
Preliminary simulations indicate that these effects are consistent with the 
total-rate behavior discussed above, and can further improve the discrimination 
power, especially when beam polarization is available at ILC. 
A more detailed differential study is left for future work.

For comparison, previous sensitivity estimates at LEP and early ILC 
studies typically relied on parton-level simulations without full ISR 
and beamstrahlung effects~\cite{7,8,33}. The present analysis therefore 
represents a more realistic projection, showing that the clean di-muon 
channel retains robust sensitivity even once collider-specific effects 
are included. This improvement is particularly relevant for CEPC, whose 
very high luminosity enhances the reach well beyond earlier expectations.

Taken together, the results presented in 
Figs.~\ref{fig:sigmavsenergy}--\ref{fig:disc_reach} 
provide a comprehensive picture of the discovery potential for a 
leptophilic photon $Z_\ell$ at future $e^+e^-$ colliders. The energy 
dependence of the cross section highlights the complementary strengths 
of CEPC and ILC, with the former providing very high statistics at 
$\sqrt{s}\approx 240$~GeV and the latter extending the reach towards 
higher masses through its 250 and 500~GeV options. The variation of the 
cross section with $M_{Z_\ell}$ and $g_\ell$ illustrates how even 
moderate couplings already produce visible deviations from the SM, even 
after ISR and beamstrahlung are taken into account. Invariant-mass 
spectra confirm that narrow resonances would appear as clear peaks over 
the SM background, consistent with existing bounds from previous $e^+e^-$ 
experiments~\cite{33,36,37}. Finally, the projected 
3$\sigma$ and 5$\sigma$ contours show that CEPC can probe couplings 
down to $g_\ell \approx 10^{-3}$ in the sub-TeV region, while the ILC 
pushes the accessible range further into the TeV domain. These findings 
underline the strong complementarity of circular and linear colliders, 
and establish a solid phenomenological basis for incorporating 
leptophilic gauge bosons into future precision programs~\cite{11,12,25,26}.

\section{Conclusions}

We have presented a dedicated study of the discovery potential for a leptophilic 
gauge boson $Z_\ell$ in the clean channel $e^+e^- \to \mu^+\mu^-$ at future 
$e^+e^-$ colliders. Incorporating the effects of initial-state radiation (ISR), 
beamstrahlung (BS), and basic detector acceptance, we have provided realistic sensitivity 
estimates for CEPC and ILC. The results show that CEPC, with its very high luminosity 
at $\sqrt{s}=240$~GeV, offers broad coverage in the low-mass region, probing 
couplings down to $g_\ell \approx 10^{-3}$, while the ILC extends the accessible 
$Z_\ell$ mass range up to the TeV scale through its 250 and 500~GeV stages. 
This complementarity underscores the unique role of next-generation lepton colliders 
in testing leptophilic gauge interactions with high precision.

Our analysis establishes a clear benchmark for the cleanest leptonic final state and 
demonstrates that even after including realistic beam effects the $Z_\ell$ signal 
remains clearly visible above the SM background. These findings provide a solid 
phenomenological basis for future strategies in exploring purely leptophilic scenarios 
at high-energy colliders.

Compared with previous parton-level estimates and earlier sensitivity projections 
from LEP and ILC studies, the present work offers a more realistic evaluation by 
incorporating ISR, BS, and acceptance effects. In this respect, our results 
represent the first direct side-by-side comparison of CEPC and ILC in the context of a 
purely leptophilic gauge boson. The projections presented here highlight the advantage 
of CEPC in probing very small couplings at low mass, and the complementary role of ILC 
in extending the discovery reach well into the multi-hundred GeV domain.

From a broader perspective, the analysis provides timely input to the global discussion 
on the physics potential of future lepton colliders. In particular, CEPC, as a 
central project in China, is shown to deliver world-leading sensitivity to new leptonic 
forces in the sub-TeV regime, while the ILC offers a complementary path at higher 
energies. Taken together, these findings underline the importance of pursuing both 
circular and linear collider options and establish a strong phenomenological basis 
for including leptophilic gauge bosons in the physics case of next-generation facilities.

\section{Data Availability}
All data generated and analyzed in this study are included in the article. Additional simulation inputs and numerical results are available from the corresponding author upon reasonable request.

\FloatBarrier
\bibliographystyle{iopart-num} 
\bibliography{refs}

\end{document}